\documentclass[%
 reprint,
 amsmath,amssymb,
 aps]{revtex4-1}

\usepackage[pdftex]{graphicx}
\usepackage[colorlinks=true,linkcolor=black, citecolor=black,
urlcolor=black]{hyperref} 

\usepackage{mathrsfs}
\usepackage[usenames,dvipsnames]{xcolor}
\usepackage{color}
\usepackage{graphicx}
\usepackage{dcolumn}
\usepackage{bm}
\def\O{\Omega}
\def\defeq{\stackrel{\text{def}}=}
\newcommand\re[1]{({\ref{#1}})} \def\be{\begin{eqnarray} }
\def\ee{\end{eqnarray}}  

    \def\no{\nonumber} \def\la{\label} 
\def\({\left(} \def\){\right)} \def\<{\left\langle\,} \def\>{\,
\right\rangle} \def\[{\left[} \def\]{\right]} 
\def\hf{ {\textstyle{1\over 2}} } 
   \def\CO{{
\mathcal{ O} }}  
 
\def\o{\omega}
       \def\CN{{ \cal N}}
       \def\CU{{\cal U}}
      \def\CV{{\cal V}}
       
      \def\a{\alpha}  
        
     \def\th{\theta} \def\G{\Gamma} 
     \def\Tr{{\rm Tr}}
   \def\Li{ \text{Li}_2}
  
\def\sla{\slash\!\! \!}
\def\zz{ { { \bf  z} }}
\def\uu{ { {\bf u} }}
\def\vv{ { \bf v}}
\def\ww{ { \bf w }}
\def\thth{ { \bm {\th } }}
\def\vv { { \bf v  }}
  \def\k{\kappa}
  
\def\llangle{  \langle\! \!\langle  }
\def\rrangle{ \rangle\!\!\rangle }
\def\vvert{ |\!\! |  }

\newcommand{\caS}{{\mathscr S}}
\newcommand{\caA}{{\mathscr A}}

\newcommand{\caZ}{{\mathscr Z}}

\begin{document}

\preprint{IPhT/t12/023}

\title{Classical Limit of the Three-Point Function from  Integrability
  } 

\author{ Ivan Kostov}
  
\affiliation{%
	     Institut de Physique Th\'eorique, CNRS-URA 2306,
	     C.E.A.-Saclay, F-91191 Gif-sur-Yvette, France }

\begin{abstract}
We give analytic expression for the three-point function of three
large classical non-BPS operators $\mathcal{N}=4$ Super-Yang-Mills
theory at weak coupling.  We restrict ourselves to operators belonging
to an $su(2)$ sector of the theory.  In order to carry out the
calculation we derive, by unveiling a hidden factorization property,
the thermodynamical limit of Slavnov's determinant.
\end{abstract}

\pacs{Valid PACS appear here}
\maketitle

\section{\label{sec:intro}Introduction}

In the last ten years, starting with the pioneer paper by Minahan and
Zarembo \cite{Minahan:2002ve}, a vast integrable structure has been
unveiled in the $\mathcal{N}=4$ supersymmetric Yang-Mills (SYM) theory
\cite{Beisert-Rev}.  There are hopes that together with the spectrum
of states, the integrability can be used to compute the correlation
functions of the theory.  Of special interest are the correlation
functions of one-trace operators in the classical limit when the
length of the traces is very large.  Such operators are dual to
extended classical strings in the AdS$_5\times$S$^5$ background, and
knowing their correlation functions can shed light about the
interactions at strong coupling.
 
Recently, Escobedo, Gromov, Sever and Vieira \cite{EGSV, GSV}
developed Bethe-Ansatz techniques for computing the tree-level
structure coefficient $C^0_{123}$ and found an expression for the
latter in terms of scalar products of Bethe states for the XXX$_{1/2}$
spin chain.  In \cite{GSV} an elegant analytic formula was derived for
the classical limit of the structure coefficient of when one of the
operators is protected (BPS).  In this note we generalize the result
of \cite{GSV} to the case of three non-BPS classical operators.  Our
starting point will be the representation of the structure constant in
terms of Slavnov-like determinants \cite{nikitaslavnov}, proposed
recently by Foda \cite{Omar}.
     
 \section{\label{sec:3pf}3-point functions of trace operators in
 $\CN=4$ SYM}

In a $su(2)$ sector of the SYM theory, the operators are made of
two complex scalars $Z$ and $X$.  We consider the correlation function
of three single-trace operators of the type $\CO_1\sim \Tr[ Z^{L_1 -
N_1} X^{N_1} + \dots]$, $\CO_2 \sim \Tr[\bar Z^{L_2- N_2}\bar X^{N_2}+
\dots ]$, $ \CO_3 \sim \Tr[Z^{L_3- N_3}\bar X^{N_3}+ \dots]$,
 where the omitted terms are weighted products of the same
 constituents taken in different order.  The weights are chosen so
 that the operator $\CO_n$ is an eigenstates of the dilatation
 operator with dimensions $\Delta_n$.  At tree level, the structure
 coefficient is a sum over all possible ways to perform the Wick
 contractions between the scalars and their conjugates.  A non-zero
 result is obtained only if $N_1 = N_2 + N_3$ and the number of
 contractions $L_{ij}$ between operators $\CO_i$ and $\CO_j$ are $
 L_{12}= L_1-N_3$, $L_{13}= N_3$, $L_{23}= L_3-N_3$.

This problem is solved using the Algebraic Bethe Ansatz \cite{EGSV}.
In the Bethe-Ansatz approach, the operator $\CO_i$ is represented by a
$N_i$-magnon Bethe eigenstate with energy $\Delta_i$ of the
XXX$_{1/2}$ spin chain of length $L_i$ ($i=1,2,3$).  To simplify the presentation
we consider only highest-weight states, but our method is valid in
general.  Such a state is completely characterized by the rapidities
of the magnons $\uu= \{u_a\}_{a=1}^L$ and will be denoted by $\vvert
\uu\rrangle_{\!  _{L}}$.
  
It is advantageous first to deform the problem by introducing
impurities $ {\thth^{(n)} } =\{\th_j^{(n)}\}_{j=1}^{L_n}$ at the sites of the
$n$-th spin chain $(n=1,2,3)$, and  take the 
homogeneous limit $\thth^{(n)} \to 0$ at the very end.  We denote the impurities associated with
the contractions between the operators $\CO_m$ and $\CO_n$ by
$\thth^{(mn)}$, so that $\thth^{(1)}= \thth^{(12)} \cup\thth^{(13)}$,
etc.  Then the  tree level structure coefficient
is given, up to a  normalization and a phase factor, by \cite{EGSV}
 \be \!\!\!\!\!\!\!  \la{defC123} {C}^{ 0}_{123} &=&
  {
  \llangle \uu \vvert \vv \!  \cup \!  \zz\rrangle_{L_1}\, 
  \llangle\zz  \vvert\, \ww \rrangle _{N_3} 
  \over 
  \llangle \uu \vvert\uu
 \rrangle_{L_1}^{1/2} \llangle \vv \vvert\vv \rrangle_{L_2}^{1/2}
 \llangle \ww \vvert\ww \rrangle_{L_3} ^{1/2}
 } \, , \ee
where $\zz= \thth^{(13)}+ i/2$ and  the r.h.s. should be evaluated in the homogeneous limit
$\zz\to\{i/2\}.$ Here the symbol $\llangle \uu \vvert\vv\rrangle_L $
stands for the scalar product of two Bethe states with rapidities $
\uu= \{u_a\}_{a=1}^N$ and $\vv=\{v_a\}_{a=1}^N$ in a spin chain of
length $L$.  In the limit when all rapidities go to infinity,
$C^0_{123}\to C^{\text{BPS}}_{123}$.

 We are interested in the {\it classical limit} $L_i \to \infty$, with
 $\a_i = N_i/L_i $
%
  finite.
As shown in \cite{Omar}, the regularization provided by the impurities
allows to express the structure constant in terms of a ratio of
determinants.  In order to obtain the classical limit of $C^0_{123}$,
we will first obtain the classical, or thermodynamical, limit of
Slavnov's determinant. 
 In our approach it is essential to evaluate  the
classical limit {\it before}  the homogeneous limit
$\zz\to i/2$.

\section{Slavnov's determinant}

{\it 1. Slavnov's formula for the scalar product. }

\smallskip

Assume that the length-$L$ $N$-magnon state with rapidities $\uu=
\{u_a\}_{a=1}^N$ a Bethe eigenstate.  Then the rapidities $\uu$
satisfy the Bethe equations, which depend on a set of impurities
$\thth=\{\th_j\}_{j=1}^L$.  The Bethe equations are equivalent to the
conditions
\be \la{Betheeq} e^{2i p_{ \uu} (z)}=-1\quad \text{for}\quad z\in\uu, \ee
where the quasi-momentum $p_\uu$ is defined as
 \be \la{defZ} 
e^{ 2ip_{{\uu}}(z)}\ \defeq \ \k \, {Q_\thth ( z-{ i\over 2})\over
Q_\thth(z+{i\over 2})}\, {Q_{\uu}(z +i)\over Q_{\uu }(z-i)}.  \ee
Here $Q_{\uu}$ and $Q_{\thth}$ are Baxter's polynomials  
\be
 \la{defQQQ} 
Q_{\uu}(z) = \prod_{ a=1 }^N (z-u_a), \ \ \ 
Q_{\thth}(z) = \prod_{ j=1 }^L (z-\th_j). 
 \ee
We also introduced a twist $\k$, which does not spoil the
integrability and  allows to handle better the singularities.
With this assumption, the scalar product $\llangle \uu \vvert\vv
\rrangle_L$ with an arbitrary Bethe state with rapidities $\vv =
\{v_a\}_{a=1}^N$ is evaluated,  in certain normalization,  by \cite{nikitaslavnov}
\be \la{Slavnovdet} 
\llangle \uu \vvert\vv \rrangle_L 
= \caS_{\uu, \vv }\,  
\defeq  \, {\det_{ab} \O_\k(u_a, v_b) \over
\det_{ab}{1\over u_a-v_b+i} }, \la{defSN}
\ee \, 
%
\be \la{defOhat} \O(u,v)= {i\over u-v}\( {1\over u-v+i} - {e^{2i p_{
\uu} (v)} \over u-v-i} \) \!  . \ee

An important particular case is the Gaudin-Izergin determinant, which gives the partition function of the 6-vertex
model with domain-wall boundary conditions \cite{Izergin-det,korepin-DWBC}, and
which we denote by $\caZ_{\uu,  \zz }$.  Gaudin-Izergin
 determinant is equal  to
$\caS _{\uu,\vv }$ with $N=L$, with the second set of rapidities frozen to  $\vv =  \thth + i/2 \equiv \zz$.  Since $Q_\thth (v)=0$
if $v - i/2 \in\thth$, the condition $\vv =\zz$ is equivalent to retaining
only the first term in the definition \re{defOhat}.  For any two sets
$\uu$ and $\vv $, not necessarily satisfying Bethe equations,
we define
\be \caZ_ {\uu, \zz}= {\det_{ab} [  {i\over (u_a-v_b)(u_a-v_b+i)} ] \over \det{1\over
u_a-v_b+i}}.  \ee


   {\it 2. Factorization property of Slavnov's determinant.} 
   
   \smallskip
   
 We will use an operator representation of Slavnov's determinant
 \re{Slavnovdet},  which we  call  {\it factorization formula}, because
 in the limit $N\to\infty$ it  factorizes into a product of 
 two  computable
 functionals.   

  \noindent $\bullet$ {\it Factorization formula: } If $\uu \cap \vv
  =0$, Slavnov's determinant \re{defSN} is given by the  
  expectation value
%
  \be \!\!\!\!\!\!  \la{SLhatb} \caS_{\uu, \vv } &=&\!\!  
  (-1)^N \
  {\langle \vv |\, \caA^+_\vv [\CU]\,\, \caA^-_\uu[\CV] |\uu \rangle \over
  \langle \vv|\uu\rangle} \, , \ee
where the functionals $\caA^\pm[f]$ are defined by
 \be \la{defCA} \caA^\pm _\uu[f] &\defeq & { \det_{ab} \( u_a^{b-1} -
 f(u_a) \, (u_a \pm i)^{b-1}\) \over \det_{ab} \( u_a^{b-1}\) }\, ,
 \ee
 and the functional arguments $\CU,\CV$ satisfy the algebra
%
%
\be \CU(z) \CV(w) =\CV(w) \, \CU(z) \, \(1 -{1\over  (z - w)^{2}+1}\)\,
 \ee
and act on the left and right  vacuum states as
 %
   \be\la{defef} \CU(v)\,\,  |\uu \rangle &=& e^{2i p_\uu(v)}
    {Q_\uu(v-i)/ Q_\uu(v)} \ |\uu \rangle \, , \la{defOplus} \no \\
   \langle \vv| \, \,  \CV(u) &=& {Q_\vv (u+i )/ Q_\vv (u)} \ \langle \vv| .
   \la{defOm} \ee

The proof of the factorization formula \re{SLhatb} will be presented
elsewhere \footnote{I. Kostov, to appear}.  Note that while the r.h.s.
of \re{defSN} makes sense only if the sets of rapidities $\uu$ and
$\vv $ have the same cardinality, the r.h.s. of \re{SLhatb} is defined
for {\it any } two sets $\{u_a\}_{a=1}^{N_1}$ and
$\{v_b\}_{b=1}^{N_2}$.

The Gaudin-Izergin  determinant 
 is evaluated by eq. \re{SLhatb}
 with $\CU=0$.
 Then  $\CV(u)$ can be 
 treated as a c-number function $V(u)$ 
 and  eq.   \re{SLhatb} becomes
\be \la{izerg} \caZ_ {\uu, \zz} & =&
(-1)^N\    \caA^-_\uu[V],
\quad V(u)  =  {Q_\zz(u+i)\over Q_\zz(u) } .  \ee

\noindent
{\it  3.  Properties of the functionals $\caA^\pm _\uu[f]$}
 
 \smallskip

The functionals $\caA ^\pm _\uu[f]$ are symmetric polynomials of
$f(u_a)$ of degree $N$, and can be expressed in terms of a sum over all possible partitions of the set $\uu$ into two subsets $\a$ and
$\bar\a$, 
\be \la{cluster} \!\!\!\!\!\!\!\!\!  \caA ^\pm _\uu [f]= \!\!\!\!\!\!
\sum_{\a\cup \bar\a = \uu }\!\!  \!(-1)^{|\a|} \prod_{a\in\a } f(u_a)
\!\!  \prod _{a\in \a, b\in \bar \a }{ u_a - u_b \pm i \over u_a-u_b},
\ee
  with $|\a|$ standing for the number of elements of the subset $\a$.
 This  expansion gives an alternative definition of $\caA^\pm_\uu$,
 which was used in \cite{GSV}  in the particular case $f(u)
=\k\, ({u- i/2\over u+i/2})^L$.

Using the expansion \re{cluster}, one can easily prove the
functional relations
\be \la{funceqa} \caA  ^\pm _{\uu}[1/f]\ \prod_{j=1}^N f(u_j)\, =(-1)^N
\, \caA ^\mp _{\uu}[f] .  \ee
 %

\section{\label{sec:SLcl} Classical  limit  of    the scalar product
of Bethe states }

\medskip \noindent {\it 3.  Classical limit of $\caA ^\pm [f]$}

\smallskip

We are interested in the classical limit $N\to\infty$, where the
points of the set $\uu$ condense into a set of contours 
cuts   $\G_\uu= \cup_k \G_\uu^k$ with linear
density $\rho(u)$.  We do not renormalize the $u$'s, so that $\rho\sim
1, u_a\sim N$.  The distribution is characterized by the resolvent
\be \la{defresolv} G_\uu(z)= \sum _{j=1}^N {1\over z- u_j} \ \simeq\
\int\limits _{\G_\uu} du\, {\rho(u) \over z-u}.  \ee

It is easy to see that  the linear term in $f$  in
\re{cluster} can be written as a contour integral,
\be \la{I1} \caA ^\pm _\uu[ f] 
&=&1 \pm \oint\limits_{C_\uu} {dz\over 2\pi }
\ f(z) {Q_\uu(z\pm i)\over Q_\uu(z)} + O(f^2)
\no \\
   & \simeq& 1\pm \oint\limits_{C_\uu} {dz\over 2\pi } \, e^{ i q_\pm(z)}+
   O(f^2), \ee
where the integration contour $C_\uu$ encircles $\G_\uu$ anticlockwise   and the function $q(z)$ is defined as
\be \la{defq} q_\pm (z) =-i \log [ f(z)] \pm G_\uu(z).  \ee
By the functional relations \re{funceqa}, similar representation 
holds  for  $f$ large. For the complete solution we
try an ansatz of the form
%
\be \la{cllA} \caA ^\pm_{_\uu}[ f] = \exp\big[ \oint\limits_{C_\uu} {dz\over
2\pi }\ F^\pm ( e^{iq_\pm (z)}) \big]\, , \ee
%
%
where the functions $F^\pm$ can be expanded as 
\be 
\la{TaylorF} 
F^\pm (\o) = F_1^\pm \o+ F_2^\pm \o^2 + F_3^\pm \o^3 + \dots ,
\ee
with $F^\pm _1 = \pm 1$.  The coefficients $F_n$ can be determined by
comparing with the exactly solvable case  $f(z)=\k$,
 or $q_\pm (z) =
-i\log \k \pm G_\uu(z)$,
where  \cite{GSV} 
\be \la{triviale} \!\!\!  \caA _{_\uu}^\pm [\k ]= (1-\k)^{ N} 
.  
\ee
To compare with \re{cllA}, we perform the contour integration  
using the asymptotics  $ e^{i q_\pm (z) } \simeq ( 1 \pm
\k\,  {N\over z}) $ at $z\to\infty$, and find $F^\pm _n= \pm 1/n^2$.  Therefore
  \be \la{defFli} F^\pm (z) &=&\pm \sum_{n=1}^\infty \ {z^n\over n^2}
  =\pm\, \Li(z) .  \ee
The functional equation for the dilogarithm, 
\be 
\la{Lifunceq}
 \Li({1/z} )= - \Li(z) - {\pi^2/6} - \hf \log^2(-z), 
\ee
is  the scaling
limit of \re{funceqa}.
 \begin{figure}[t]
\includegraphics[scale=.44]{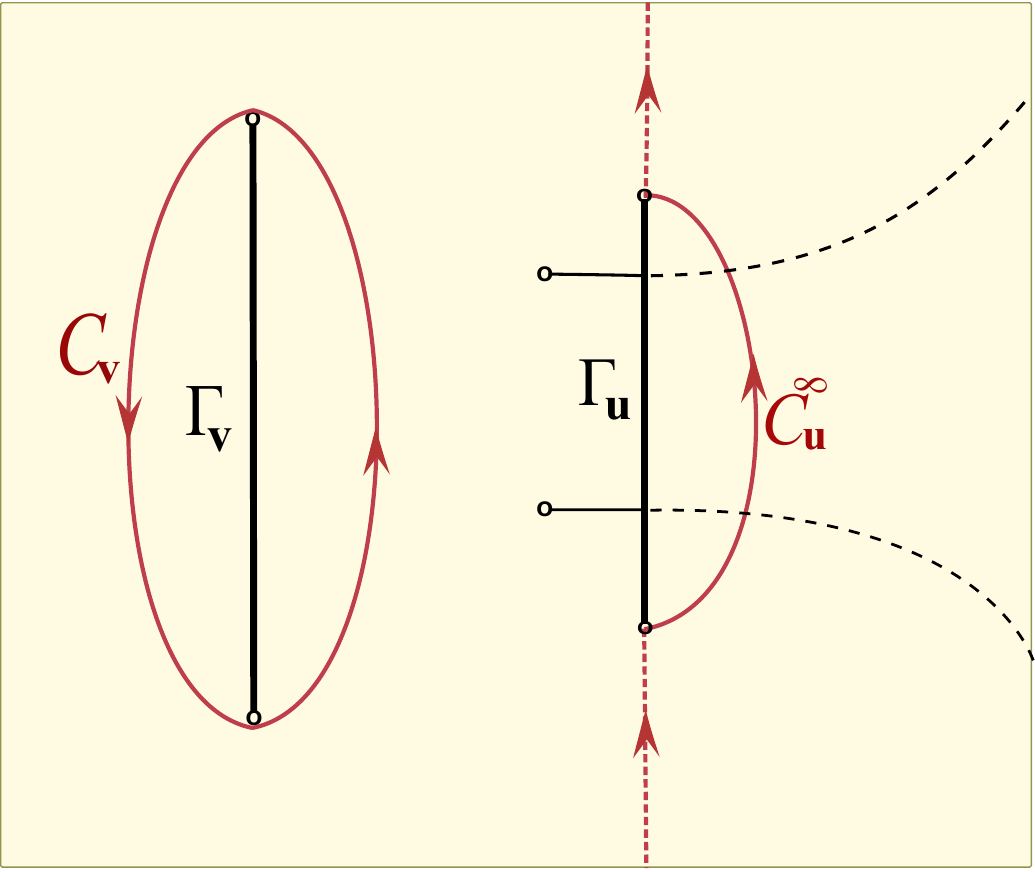} 
\hskip 0.8cm 
\includegraphics[scale=.58]{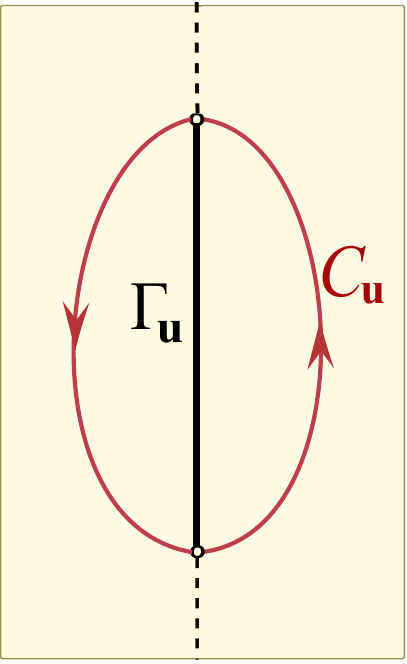} 
\caption{\label{fig:Contour} 
Left: The contour $C_\vv $ and deformed contour $ C_\uu^\infty$ for
the integral in \re{intrepf} in the case when $\G_\uu$ and $\G_\vv$
have one connected component.
 Right: The contour $C_\uu$ for the integral in \re{classnorm}.  When
 $G_\vv\to G_\uu$, the two logarithmic branch points on the first
 sheet join the two simple branch points at the extremities of
 $\G_\uu$.  }
\end{figure}

\bigskip

 {\it 4.  Classical limit of the Slavnov and Gaudin-Izergin determinants and
 of the Gaudin norm }
 
\smallskip

We will use the factorization formula \re{SLhatb} to find for the
classical limit of the Slavnov determinant \re{defSN}.  In this limit
we can consider $\CU$ and $\CV$ as $c$-number functions, since the they
commute up to $O(N^{-2})$.  Then we can use the functional relation
\re{funceqa} to write \re{SLhatb} in the form
    \be \!\!\!\!\!\!  \la{SLclass} 
    \caS_{\uu,\vv } &=&\!\!  \,\caA ^+_\vv
    [\k\, e^{i G_\uu - i G_\thth]} ]\,\, \caA ^-_\uu[e^{i G_\vv}] \,
     .\ee
%

Introduce, as in \re{defresolv}, the resolvents $G_\uu, G_\vv $ and
$G_\zz$, associated respectively with the sets of points $\uu,\vv $
and $\zz$.  The  classical limit of Slavnov's scalar product is
obtained by substituting \re{cllA} in the factorization formula
\re{SLclass}:
  \be \la{intexplnreg} \hskip -0.7cm \log \caS _{\uu,\vv }\!  =\!  \oint
  \limits_ {C_\vv} \!  \frac{dz}{2\pi } \ \text{Li}_2 ( e^{i\,
  q})\!  - \oint \limits_{C_\uu }\!  \frac{dz}{2\pi } \ \text{Li}_2(
  e^{ i G_ \vv- iG_\uu}) , \ee
%
%
  \be \la{defqph} q&\defeq & G_\uu+ G_\vv - G_\thth
  +\log\k.  \ee
The integration contours $C_\uu$ and $ C_\vv$ encircle $\G_\uu$ and
$\G_\vv $ anticlockwise.  

The r.h.s. of \re{intexplnreg} can be
reformulated entirely in terms of the function $q(z)$ defined in
\re{defqph}.   
The Bethe equations  \re{Betheeq} imply 
a boundary condition for the resolvent  $G_\uu$, 
\be 2 \sla G_\uu(z ) - G_\thth(z) +\log \k = 2\pi n_k \ \ \text{for} \
\ z\in \G_\uu^{k}, \ee
where $\sla G_\uu$ is the half-sum of the values of the resolvent on
both sides of $\G_\uu$ and $n_k$ is the mode number associated with
the $k$-th connected component $\G^k_\uu\subset \G_\uu$.  Hence, if
$q^{(1)} $ is the value of the function $q(z)$ on the physical sheet
defined by \re{defqph}, then the value of $q(z)$ on the second sheet
is given by $q^{(2)} = - G_\uu+ G_\vv $ and \re{intexplnreg} can be
written as
\be \la{intrepf} \log \caS _{\uu,\vv }&=& \oint \limits_{C _\uu \cup
C_\vv } \frac{dz}{2\pi } \ \text{Li}_2(e^{i\, q(z)}).  \ee
(The minus sign is compensated by the change of the orientation of
contour $C_\uu$  after it is moved to the first
sheet.)  The integral along $C_\uu$ is however ambiguous, because the
integrand has two logarithmic cuts which start at two branch points on
the first sheet and end at $z=\infty$ on the second sheet, after
crossing the cut of the resolvent $G_\uu$ on $\G_\uu $.  The
ambiguity is resolved by deforming the contour $C_\uu $ to a contour
$C^\infty_\uu$ which encircles also the point $z=\infty$ on the second
sheet \footnote{The
author is indebted to Nikolay Gromov for performing the numerical test
and for suggesting how to place the integration contours.}.  In the case of a one-cut solution, the contour $C_\uu ^\infty$
is depicted in Fig.  \ref{fig:Contour}, left.  With this prescription,
eq.  \re{intexplnreg} reproduces the numerical data (for $\k=-1$ and
$N$ up to 60) with precision $10^{-12}$.  Another test of \re{intrepf}
is to send all the roots $\uu$ to infinity.  In this limit the
integration goes only along the contour $C_\vv$ and the function $q$
in the integrand is given by $q= G_\vv- {1\over 2} G_\thth $.  Then eq. \re{intrepf} reproduces correctly the expression  obtained 
in  \cite{GSV} for the scalar product of  Bethe
state and a vacuum descendent.  

We will also need the classical limit of the Gaudin-Izergin determinant, for
which \re{izerg} gives
  \be \la{clsIzer} && \log\caZ_ {\uu,\vv }
=
     - \oint \limits_{C_\uu } \frac{dz}{2\pi } \ \text{Li}_2\left( e^{
     i G_\vv -iG_\uu}\) 
 .
\ee

Finally, an expression for the square of the Gaudin norm can be
formally obtained from \re{intrepf} by taking $G_\uu=G_\vv =G$.  When
$\G_\vv\to \G_\uu$, the integration contour in \re{intrepf} can be
closed around $\G_\uu=\G_\vv$ as in Fig.  \ref{fig:Contour},
right, and $q$ in the integrand  is  replaced
 by $ 2p_\uu$, where  
\be 
p_\uu &=& G_\uu - \hf G_\thth + \hf \log k 
\ee
 is the quasi momentum.  Thus we find for the square of the Gaudin
 norm
 \be \la{classnorm} \log \caS _{\uu, \uu } &=& \oint \limits_{ C_{\uu} }
 \frac{dz}{2\pi } \ \Li \( e^{ 2i p_\uu(z)}\).  \ee
  %
One can check, using the fact that $p(z) = \pm i\pi
\rho(z)$ on the two edges of the cut, that the contour integral
\re{classnorm} can be transformed into (twice) the linear integral in
eq.  (2.15) of \cite{GSV}.

\section{\label{sec:3pcl}Classical limit of the structure constant}

 Now we can proceed with the computation of the 
 classical limit of the structure constant  \re{defC123},
 which we express in terms of the functionals considered 
 above,
\be\!\!\!\!  \la{C123final} 
{ {C}^{ 0}_{123} } = {\caS _{\uu, \vv\cup\zz}\ \
\caZ_ { \zz, \ww}  
\over \caS _{\uu,\uu}^{1/2}\ \ \caS _{\vv,\vv}^{1/2}\ \
\caS _{\ww,\ww}^{1/2} }.  \ee
    In applying \re{intrepf}, \re{clsIzer} and \re{classnorm}
    the only
 non-obvious point is the evaluation of $\caS _{\uu , \vv \cup
 \zz}$ with $\zz = \thth ^{(13)} +{i\over 2}$.  This is the the
 `restricted Slavnov product' studied in \cite{1999NuPhB.554..647K,
 2011NuPhB.852..468W, Omar}, in which part of the magnon rapidities
 are frozen to the values of the impurities on a segment of the spin
 chain.  In the original formulation \re{Slavnovdet}, the restricted
 Slavnov product is given by a ratio of vanishing quantities, which
 necessitates to apply repeatedly l'H\^opital's rule.
  In contrast, the  factorized representation \re{SLhatb} 
   is free of  such complications.  
   It  is given by the r.h.s. of 
 \re{intrepf}, with    (for  $\k=1$) 
 \be
 \la{defqrestr}
  q &= & G_{\uu}+ G_{\vv\cup \zz} - G_{\thth^{(12)} \cup \thth^{(13)}}
  \no
  \\
  &= & G_\uu + G_{\vv}- G_{\thth ^{(12)} }
 .
   \ee
Expressing the resolvents $G_\uu,G_\vv,G_\ww$ 
in terms of the 
 three quasi-momenta
 $p_\uu = G_\uu - \hf G_\thth^{(1)}$, 
 $p_\vv = G_\vv - \hf G_\thth^{(2)}$ and 
$p_\ww = G_\ww - \hf G_\thth^{(3)}$, 
and taking the 
  homogeneous limit
 $\thth^{(n)} \to 0$,  replacing 
 $G_{\thth^{(n)}}\to {L_n/2z}$ ($n=1,2,3$).
 we finally obtain, up to a complex constant,
\be \log \la{clasC123} { C^0_{123}   } &\simeq&\ -\!\!\!
\sum_{n= \uu, \vv, \ww} \hf \oint \limits_{ C_n} \frac{dz}{2\pi } \ \Li
\big[e^{ 2i p_n(z)}\big] \no \\
&+& \!\!\!  \oint \limits_{ C^\infty_\uu\cup C_\vv } \!\!\!\!\!\!\!\!
\frac{dz}{2\pi } \ \text{Li}_2\big[e^{ i p_{\uu}(z) + i p_\vv(z)+i {
L_3/2z}} \big] \no \\
  &+& \oint \limits_{C_\ww} \frac{dz}{2\pi } \text{Li}_2\left[ e^{ i
  p_\ww(z)+i { (L_2- L_1)/2 z} } \right]\!  . 
  \ee
%
%
%
As it was pointed out by
Gromov and Vieira in \cite{2012arXiv1202.4103G}, the tree level solution
for $C^0_{123}$ in presence of impurities 
$\thth^{(1)}, \thth^{(2)}, \thth^{(3)}$  can be used to obtain the
one-loop corrections. In this sense  we have obtained
also the  correlator of three non-BPS classical fields at one-loop. 

 The method outlined in this note allows to
handle the impurities in the classical limit and attack the problem in
its full generality.  The expression \re{clasC123}  can be used
\footnote{D. Serban, to appear.} to show that, at least in the
classical limit, the two-loop result is obtained by changing the
quasimomenta $p_\uu, p_\vv$ and $p_\ww $ according to the three-loop Bethe
ansatz equations \cite{Beisert:2004hm}. 
  It is natural to expect  that the full structure coefficient in
the $SU(2)$ sector in SYM will be obtained from \re{clasC123} by using
the exact expression for the quasimomenta upon inclusion of the
dressing phase \cite{Beisert:2006ez}.
  At least this possibility is worth of being explored  and we hope to be able to report on this  in a future publication.

  \begin{acknowledgments}
   The author is obliged to  S. Alexandrov, O. Foda, N. Gromov, A. Sever, D.
   Serban, P. Vieira and K. Zarembo for illuminating discussions,  
    to P. Vieira and N. Gromov for conducting the numerical tests, and
  to  P. Vieira and O. Foda for critical reading of the manuscript.  
   Part of this work has been
   done during the visit of the author at Nordita in February 2012.
\end{acknowledgments}

%
 
%

\begin{thebibliography}{10}

\bibitem{Minahan:2002ve}
J.~A. Minahan and K.~Zarembo,
{\em JHEP} {\bf 03} (2003) 013,
\href{http://arXiv.org/abs/hep-th/0212208}{{\tt
arXiv:hep-th/0212208}}.

\bibitem{Beisert-Rev}
  N.~{Beisert {\it et al}}, ``{Review of AdS/CFT Integrability: An
  Overview},'' {\em Lett.  Math.  Phys.} {\bf 99} (Jan., 2012) 3--32,
  \href{http://arXiv.org/abs/1012.3982}{{\tt arXiv:hepth/1012.3982}}.

\bibitem{EGSV}
J.~Escobedo, N.~Gromov, A.~Sever, and P.~Vieira, 
   \href{http://arXiv.org/abs/1012.2475}{{\tt
   arXiv:hep-th/1012.2475}}.

\bibitem{GSV}
N.~{Gromov}, A.~{Sever}, and P.~{Vieira},
    \href{http://arXiv.org/abs/1111.2349}{{ \tt
    arXiv:hep-th/1111.2349}}.

\bibitem{nikitaslavnov}
N.~A. Slavnov, 
  {\em Russian Math. Surveys} {\bf 62}:4 (2007),
   727.


\bibitem{Omar}
O.~{Foda}, 
  \href{http://arXiv.org/abs/1111.4663}{{\tt arXiv:hep-th/1111.4663}}.


\bibitem{Izergin-det}
A.~G. {Izergin},
  {\em Soviet Phys.  Doklady} {\bf 32} (Nov., 1987) 878.

\bibitem{korepin-DWBC}
V.~E. Korepin,
 {\em
  Comm. Math. Phys.} {\bf 86} (1982) 391--418.

\bibitem{Note1}
I. Kostov, to appear.

\bibitem{Note2}
The author is indebted to Nikolay Gromov for performing the numerical
test and for suggesting how to place the integration contours.


\bibitem{1999NuPhB.554..647K}
N.~{Kitanine}, J.~M. {Maillet}, and V.~{Terras}, 
  {\em Nucl. Phys.  B} {\bf 554} (1999) 647,
    \href{http://arXiv.org/abs/arXiv:math-ph/9807020}{{\tt
  arXiv:math-ph/9807020}}


\bibitem{2011NuPhB.852..468W}
M.~{Wheeler},
   {\em Nucl. Phys. B} {\bf 852} (2011)468, 
   \href{http://arXiv.org/abs/1104.2113}{{\tt arXiv:math-ph/1104.2113}}.


\bibitem{2012arXiv1202.4103G}
N.~{Gromov} and P.~{Vieira},
  \href{http://arXiv.org/abs/1202.4103}{{\tt arXiv:hep-th/1202.4103}}.

\bibitem{Note3}
D. Serban,  \href{http://arxiv.org/abs/1203.5842}{{\tt arXiv:hep-th/1203.5842}}

\bibitem{Beisert:2004hm}
N.~Beisert, V.~Dippel, and M.~Staudacher, 
   {\em JHEP} {\bf 07} (2004) 075,
   \href{http://arXiv.org/abs/hep-th/0405001}{{\tt
   arXiv:hep-th/0405001}}.

\bibitem{Beisert:2006ez}
N.~Beisert, B.~Eden, and M.~Staudacher, 
  {\em J. Stat.  Mech.} {\bf 0701} (2007) P021,
  \href{http://arXiv.org/abs/hep-th/0610251}{{\tt
  arXiv:hep-th/0610251}}.

\end{thebibliography}
   \bibliographystyle{/Users/vani/Files/PAPERS/PAPERSLIBRARY/utcaps}

 
 \providecommand{\href}[2]{#2}\begingroup\raggedright
\endgroup
\end{document}